\documentclass[prb,preprint]{revtex4-1} 

\usepackage{amsmath}  
\usepackage{amsfonts} 
\usepackage{graphicx} 
\usepackage{float}

\usepackage{color}
\newcommand\bb[1] {   \mbox{\boldmath{$#1$}}  }

\def\gtsima{$\; \buildrel > \over \sim \;$}
\def\gtsim{\lower.5ex\hbox{\gtsima}}

\newcommand\dd{\partial}

\newcommand\beq{ \begin{equation} }
\newcommand\eeq{ \end{equation} }

\begin{document}
\hfill {\it Submitted to the American Journal of Physics.}

\title{Determining the difference between local acceleration and local gravity: applications of the equivalence principle to relativistic trajectories}

\author{Steven A.\ Balbus}
\email{steven.balbus@physics.ox.ac.uk} 

\affiliation{Department of Physics, Astrophysics, University of Oxford, Denys Wilkinson Building, Keble Road, Oxford OX1 3RH, United Kingdom}



\date{\today}

\begin{abstract}
We show by direct calculation that the common Equivalence Principle explanation for why gravity must deflect light is quantitatively incorrect by a factor of three in Schwarzschild geometry.   It is therefore possible, at least as a matter of principle, to tell the difference between local acceleration and a true gravitational field 
by measuring the local deflection of light.  
We calculate as well the deflection of test particles of arbitrary energy, and construct a leading-order coordinate transformation from Schwarzschild to local inertial coordinates, which shows explicitly how the effects of spatial curvature manifest locally for relativistic trajectories of both finite and vanishing rest mass particles.  
\end{abstract}

\maketitle 

\section{Introduction}


An argument often put forth to show that gravity must affect the propagation of light proceeds along the following lines.   Consider an elevator accelerating in a designated ``upward'' direction in otherwise empty at a constant rate $g$.   Through a hole in one wall a light source emits a photon horizontally, at a right angle to the direction of acceleration.   On the opposite facing wall, a distance $H$ from the emitting wall, there is a hole for the photon to exit.   
In the accelerating frame of an observer Alice in the elevator, the light emerges from the first hole and exits the second traveling in an apparent parabolic arch, a result of the elevator's uniform upward acceleration.   Of course, from the point of view of an external observer Bob, in the finite time it has taken the photon to traverse the elevator $H/c$, where $c$ is the speed of light, it is the elevator that has risen while the photon travels in a straight line.   In particular, 
the photon emerges on the far wall a vertical kinematic distance  
\beq\label{kin}
\Delta z_{\rm kin} =  - {gH^2\over 2c^2},
\eeq
below its starting coordinate, the minus sign denoting the downward deflection.  The Equivalence Principle (EP) of general relativity is often interpreted to mean that this same vertical drop $\Delta z_{\rm kin}$ would emerge in a local calculation for the deflection of light by a gravitational field.  In other words, gravity must affect the propagation of light, causing a deflection of precisely this amount.   
The argument that an upwardly accelerating frame of reference in Minkowski space precisely reproduces the deflection of light due to gravity, is made explicitly in a widely-used textbook\cite{har}: ``[F]or an observer in the accelerating laboratory the light ray falls with an acceleration $g$.   The equivalence principle implies that the same observation is made in a laboratory at rest in a uniform gravitational field.''    A yet more recent textbook\cite{sc}, part of an excellent introductory physics series, describes the deflection of light arising from an upwardly moving elevator as ``the same thing'' arising from a local gravitational field.


The purpose of this paper is to revisit this simple argument.  It is correct for an {\it exactly} uniform gravitational field, but it is, in fact, quite incorrect for, {\it e.g.,} the rocket shown in the first textbook\cite{har} at rest in the local gravitational field of planet Earth, even though this field is indeed very nearly uniform over the scales of interest here.   Therein lies the surprise.   The correction factor is not small: in the local gravitational field of a nonrotating spherical object, the correct angular deflection of light is a factor of three larger than the right side of equation (\ref{kin}).   The original argument is incorrect not just for photons, but also for massive particles when their Lorentz factors deviate measurably from unity.  A careful local application of the laws of special relativity does not ensure the indistinguishability of a freely-falling reference frame in a gravitational field and local Minkowski geometry.   Instead, careful observations would allow the two to be discerned. 

More care than is customarily given is needed when applying the EP to the orbital deflections of photons or relativistic particles.  Some of the difficulty is that there is often ambiguity or differences in the literature in the way the EP is formulated\cite{dls}.   In this paper, we will follow Weinberg's text\cite{abb}, whereby the EP means that it is always possible to find coordinates in which the first derivatives of the metric tensor vanish locally in the neighborhood of any nonsingular point of spacetime.   Inconsistencies can arise if this is interpreted kinematically, i.e.\ that we may calculate local physics by going into an inertial, freely-falling frame of reference.  Sometimes this may work, but not all the time.   To see this more clearly, in the next section (\S2), we carry out an explicit calculation of the local photon deflection angle for standard Schwarzschild geometry, {followed in \S3} by the same calculation for a test particle of arbitrary energy.   Neither of these cases can be understood by na\"ive kinematic arguments and applications of the EP.    To probe further, in \S4 we calculate an explicit coordinate transformation between standard (cartesian) Schwarzschild coordinates and a true set of local inertial coordinates about a particular point in spacetime in which the metric is locally Minkowskian.   These local inertial coordinates are spatially curved, and the key point is that this curvature becomes an order unity effect, {\em even locally,} for the very small deflection angles associated with relativistic trajectories.    Herein lies the resolution of the discrepancy.   Finally, in the concluding discussion (\S 5), we argue that deflection of relativistic or nonrelativistic particles may be very generally calculated by assuming that both a local elevator, and the spatially curved coordinate surface through which the particle follows its geodesic, are falling vertically with the {\it same} rate of acceleration.  It is in this sense that a kinematic formulation of the EP may be recovered.

\section {Local deflection of light}
We begin with a detailed calculation of what the local deviation angle is for a photon coming off its point of closest approach from a central point mass in Schwarzschild geometry.  We work in geometric units in which Newton's $G$ and the speed of light $c$ are set equal to unity, and use the sign convention $-+++$ for the spacetime metric.   In standard quasi-spherical coordinates, the exact Schwarzschild metric takes the form
\beq\label{AB}
-d\tau^2 = -B(r) dt^2 +A(r)dr^2 + r^2d\theta^2 + r^2\sin^2\theta d\phi^2,
\eeq
where 
\beq
B(r) = 1- 2r_g/r , \quad A(r) = (1-2\alpha r_g/r)^{-1},
\eeq
and $r_g$ is the gravitational radius, which in geometric units is simply the mass $M$.   A factor of $\alpha$ has been inserted to ``tag'' the spatial curvature for later reference; it is of course equal to unity for the standard Schwarzschild metric and has no other significance.    The null geodesic orbital equation may be written\cite{abb}:
\beq\label{rrr}
{1\over r^4}\left(dr\over d\phi\right)^2 +{1\over r^2 A} = {1\over J^2 AB},
\eeq
where $J$ is the angular momentum constant defined by the $\phi$ orbital equation\cite{abb}:
\beq
r^2{d\phi\over dt} = J B.
\eeq
Writing equation (\ref{rrr}) in terms of $u=1/r$, we find
\beq
(u')^2 +u^2(1-2\alpha u r_g) = {1\over J^2}\ \  {1-2\alpha u r_g \over 1-2ur_g},
\eeq
where $u'\equiv du/d\phi$.  
The parameter $r_g$ is regarded as small here to ensure that the deflection angle is likewise small, and we shall therefore retain terms only through linear order in $r_g$.   Expanding in this manner to linear order and differentiating leads to
\beq\label{ulight}
u''  + u -3\alpha u^2 r_g = {r_g(1-\alpha)\over J^2}.
\eeq
When $\alpha=1$, the right side of this equation disappears and we recover the standard photon geodesic equation for the Schwarzschild metric\cite{bur}, which in this case is exact, without the need for an $r_g$ expansion.   

The leading order behaviour of the solution to this differential equation corresponds to ignoring terms proportional to $r_g$.   As in standard treatments of light deflection, we take our zeroth order solution to be a straight line of the form
\beq
u_0 = {\sin\phi\over b}
\eeq
where $b$ is an integration constant corresponding to the point of closest approach at an angle $\phi =\pi/2$. (See Fig. [1].)   Here, we are interested in the local path $\pi/2 \le \phi \le \pi/2 +\delta$, where $\delta$ is very small.  
The angular momentum constant $J$ appears in a small $r_g$ term, and may therefore be replaced by its leading order value, $J=b$.

The departure from a straight line enters at linear order in $r_g$.   Taking $u = u_0+u_1$, with $u_1\ll u_0$, we easily find
\beq\label{8}
u_1'' + u_1 = {3\alpha r_g\over 2b^2}(1-\cos 2\phi) +{r_g(1-\alpha)\over b^2} = {r_g\over 2b^2}(\alpha+2) - {3r_g \alpha\over 2b^2}\cos 2\phi.
\eeq
(We have replaced $u$ by its leading form $u_0$ in the term $3\alpha u^2 r_g$.)
Equation (\ref{8}) is a linear differential equation, and it is easy to see that its solution is a linear superposition of a constant plus another constant times $\cos(2\phi)$.   Trying a solution of this form determines the two constants, and we find:
\beq
u_1 = {r_g\over 2b^2}(\alpha+2 +\alpha \cos 2\phi ).
\eeq
Hence
\beq
 {1\over r} = u = u_0+u_1= {\sin\phi\over b} +{r_g\over 2b^2}(\alpha+2 +\alpha \cos 2\phi ).
 \eeq
 To leading order in $r_g$,
 \beq\label{req}
 r = {b\over \sin\phi}\left[ 1 -{r_g\over 2b\sin\phi}(\alpha+2 +\alpha \cos 2\phi )\right].
 \eeq
 The point of closest approach still occurs at $\phi=\pi/2$ through first order in $r_g$.  However, its magnitude is no longer $b$ but
 \beq
 \tilde b = b -r_g,
 \eeq
 which follows directly from (\ref{req}).   Note that this is independent of $\alpha$.   
 \begin{figure}
\begin{minipage}[m] {0.95\linewidth}
\centering
\includegraphics [width = 0.95\linewidth]{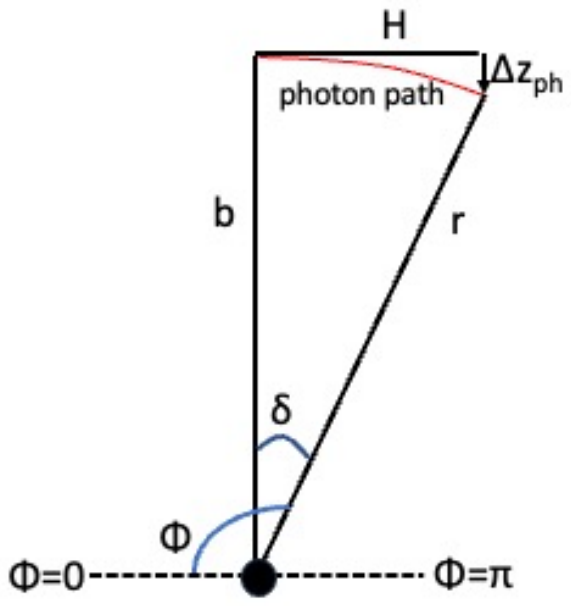}
\caption{\footnotesize Geometry for local light deflection from the initial point of closest approach $r=b$, which occurs at angular coordinate $\phi=\pi/2$.  The large black dot marks the location of the origin of the central mass.   The angle $\delta$ by which $\phi$ advances beyond $\pi/2$ is assumed to be small.   The vertical fall of the photon $\Delta z_{ph}$ is given by equation (\ref{orb}).  }
\end{minipage}
\end{figure}

 Returning to (\ref{req}), we regroup the terms so that the deviation of the photon orbit from a straight line, denoted  $\Delta z_{\rm ph}$, is explicit:
 \beq\label{line}
\Delta z_{\rm ph}\equiv r -{\tilde b\over \sin\phi} = {r_g\over \sin\phi} -{r_g\over 2\sin^2\phi}(\alpha+2+\alpha\cos 2\phi).
 \eeq
 We now expand the right side of equation (\ref{line}) about the radius of closest approach by setting $\phi =\pi/2 + \delta$, with $\delta \ll 1$.  With 
 $$
 \sin\phi =\cos\delta \simeq 1 -\delta^2/2, \qquad \cos 2\phi  = -\cos 2\delta \simeq 2\delta^ 2 -1,
 $$
 we find that (\ref{line}) gives to leading (quadratic) order in $\delta$,
 \beq\label{orb}
\Delta z_{\rm ph} =  - {r_g\delta^2\over 2}(1+ 2\alpha).
  \eeq
For a comparison with the kinematic result (\ref{kin}), we return to physical units, inserting $G$ and $c$.   With $r_g = GM/c^2$, $g=GM/r^2$ and $\delta = H/r$,     for the standard Schwarzschild geometry we find that
 \beq
 \Delta z_{\rm ph} = - {gH^2\over 2c^2}(1+2\alpha) = 3\Delta z_{\rm kin},
 \eeq
 a substantial discrepancy.   The result of the kinematic formulation of the Equivalence Principle (EP) value is recovered only by setting $\alpha =0$, which alerts us to the importance of spatial curvature.   Massive particles must behave indistinguishably from photons in the extreme relativistic limit, so it is of interest to understand how this deflection discrepancy factor arises continuously, from nonrelativistic to fully relativistic test particle motion.
 
 \section {Local deflection of a finite mass particle}
 The geodesic equation of motion for the time $t$ coordinate in Schwarzschild geometry is\cite{bur}
 \beq
 {dt\over d\tau } \equiv \dot t = {\gamma\over B},
 \eeq
where $\gamma$, the energy integration constant, is the Lorentz factor for the particle at infinity.   (For bound geodesics $\gamma <1$, and this interpretation no longer holds.)   The angular geodesic equation is 
 \beq
 r^2{d\phi\over d\tau}\equiv r^2  \dot \phi = \gamma J.
 \eeq
It will be convenient to include a factor of $\gamma$ here, which means that $J$ is the angular momentum as measured by a distant, rather than a comoving, observer.   The equation for $\dot r$ is
\beq
 -B\dot t^2 +A\dot r^2 +r^2\dot\phi^2= - 1.
 \eeq
We now restrict ourselves to the standard Schwarzschild metric.  Setting $\alpha=1$, multiplying by $(d\tau/d\phi)^2=r^4/\gamma^2J^2$, and simplifying brings us to
 \beq
 {1\over r^4}\left(dr\over d\phi\right)^2  +\left(1-{2r_g\over r}\right) \left( {1\over r^2} + {1\over \gamma^2 J^2}\right) = {1\over J^2}.
 \eeq
 As in \S 2, we set $u=1/r$ and differentiate with respect to $\phi$.   We now find
 \beq\label{full}
 u'' + u -3r_gu^2 -{r_g\over \gamma^2 J^2}=0.
 \eeq	
We recover equation (\ref{ulight}) with $\alpha=1$, and $\gamma\rightarrow \infty$.   The Newtonian limit is recovered by dropping the term $3r_g u^2$ and setting $\gamma=1$.
 
To proceed, we set $J=\beta b$, where $b$ is once again the leading order point of closest approach, and $\beta$ is the velocity at this point.  We the find:
\beq
u'' + u   =  {r_g\over \gamma^2 \beta^2 b^2} + 3r_gu^2   .
\eeq
As $\beta$ appears in a term proportional to $r_g$, it may be treated as a constant, $\beta^2 =1-1/\gamma^2$.
Indeed, the entire right side is regarded as small, since, as before, we seek only slight deviations from straight line solutions.  The leading order solution is once again $u=u_0 =\sin\phi /b$, and the correction term $u_1$ now satisfies 
\beq
u_1'' + u_1   =   {r_g\over \gamma^2 \beta^2 b^2}+ 3r_g {\sin^2\phi\over b^2}.
\eeq
To solve this equation, the approach used in \S 2 works equally well here, and yields the result:
\beq
u = {\sin\phi\over b} +{r_g\over b^2} \left[  {1\over \gamma^2 \beta^2 } + {1\over 2}(3+\cos2\phi) \right].
\eeq
To leading order in $r_g$, 
\beq
r ={1\over u} = {b\over \sin\phi} \left[  1 -{r_g\over b\beta^2 \gamma^2\sin\phi} - {r_g(3+\cos2\phi)\over 2b\sin\phi}\right].
\eeq
In terms of the ``updated'' point of closest approach,
\beq
\tilde b = b -r_g -{r_g\over \beta^2\gamma^2},
\eeq
the departure from a straight line orbit is
\beq
r -{\tilde b\over \sin\phi} = {r_g\over \sin\phi} + {r_g\over \beta^2\gamma^2} \left( {1\over \sin\phi} - {1\over \sin^2\phi}\right) -{r_g\over 2\sin^2\phi} (3+2\cos 2\phi).
\eeq
With $\delta = \phi-\pi/2\ll 1$, this becomes to leading order in $\delta$:
\beq
\Delta z_{\rm m}  \equiv r -{\tilde b\over \sin\phi}  \simeq -{r_g\delta^2 \over 2}\left( 3 +{1\over \beta^2\gamma^2} \right).
\eeq
This may also be written
\beq\label{morb}
\Delta z_{\rm m} = -{r_g\delta^2 \over 2\beta^2 }\left( 3 - {2\over \gamma^2}\right) =  -{r_g\delta^2 \over 2 }\left( {1\over \beta^2}  +2 \right),
\eeq
in which the Newtonian $\gamma = 1$ and photon $\gamma\rightarrow\infty$ limits are manifest.   For comparison, the prediction via a kinematic treatment in an upward accelerating reference frame followed by a na\"ive application of the EP would be  $\Delta z_{\rm m} = -r_g\delta^2 /2\beta^2$.   This is evidently the dominant contribution from the rightmost term in the Newtonian limit.   As $\beta$ increases however, the deflection angle differs from this EP value, decreasing in magnitude until it hits the ``photon floor'' at $\beta=1$.   We see that there is nothing unusual about photons {\it per se} deviating from na\"ive EP predictions:  the discrepancy is a generic relativistic effect, with deviations beginning to appear when the Lorentz factor of the particle is measurably different from unity.

 \section {Transformation to local inertial coordinates}
We now show how adherence to a kinematic approach to the EP reemerges when care is taken to work with proper local inertial coordinates, in which the metric appears to be locally Minkowskian.  These coordinates differ from a simple, freely-falling, frame of reference.    They are closely related to Riemann normal coordinates and have an extensive literature (see e.g.\ the text of Misner, Thorne, and Wheeler\cite{mtw}).   In this section we give explicit formulae for going between standard Schwarzschild coordinates and a set of Riemann normal coordinates.  

The Schwarzschild metric in quasi-Cartesian $(X,Y,Z)$ spatial coordinates is given by\cite{abb}:
 \beq\label{met}
 -d\tau^2 = \eta_{\alpha\beta} dX^\alpha dX^\beta  +\left(2r_g\over r\right) dt^2 + \left[2r_g\over r^2(r-2r_g)\right](\bb{X\cdot dX})^2,
 \eeq
where $\bb{X} = (X, Y, Z)$ is treated as a Cartesian position vector in the dot product, and $dX^\alpha$ is a 4-vector, with $dX^t \equiv dt$.    Here, $r^2=X^2+Y^2+Z^2$.   We shall work in a small neighbourhood of the fixed point $\bb{r_0} = (0,0,r_0)$ by defining local spatial coordinates $(x,y,z)$ such that the position vector $\bb{X}$ takes the form
 \beq
 \bb{X} =(X, Y, Z) = (0,0,r_0) + (x,y,z) \equiv \bb{r_0} +\bb{x}
 \eeq
with $x, y, z$ and $r_g$ all now assumed to be $\ll r_0$.   The time coordinate $t$ is unconstrained.   Including the first order correction term in the coordinates, 
 \beq
 (\bb{X\cdot dX})^2 = r_0^2 dz^2 +2r_0 dz (\bb{x\cdot dx}).
\eeq
 
 Next, we expand the metric (\ref{met}) to linear order in $r_g$ in the spatial coordinates $x^i$.  
 After retaining terms through order $r_g \bb{x}/r_0^2$ in the metric, we find 
 \beq\label{19}
 -d\tau^2 = \eta_{\alpha\beta} dx^\alpha dx^\beta  +\left(2r_g\over r_0\right) \left[
 \left(1 -{z\over r_0}\right) (dt^2 +dz^2)  +{2dz \over r_0} (xdx+ydy)\right].
 \eeq
(Note that for applications at the surface of the earth, $r_g$ is 4.4 mm, $r_0$ the radius of the earth, and $|\bb{x}|$ a particle travel distance, typically intermediate in scale between the two.)   
 
 Our task now is to construct local inertial coordinates  $x'^\alpha$ in which this metric by definition becomes
 \beq
 -d\tau^2 = \eta_{\alpha\beta} dx'^\alpha dx'^\beta
 \eeq
 through order $(x')^2$.   The transformation takes the functional form
 \beq
 x'^\alpha  = x^\alpha +\xi^\alpha(x),
 \eeq
 where the $\xi^\alpha$ are functions to be solved for.   
 We shall once again work to linear order in $\xi^\alpha$, and use the index convention $\xi_\alpha = \eta_{\alpha\beta} \xi^\beta$.   Then:
 \begin{align}
   -d\tau^2 &  = \eta_{\alpha\beta} dx'^\alpha dx'^\beta\nonumber\\
   & = \eta_{\alpha\beta} \left( dx^\alpha  +{\dd\xi^\alpha\over \dd x^\sigma}dx^\sigma\right) \left( dx^\beta +{\dd\xi^\beta\over \dd x^\rho} dx^\rho\right)\nonumber\\
   &\simeq \eta_{\alpha\beta} dx^\alpha dx ^\beta + 2\eta_{\alpha\beta} {\dd\xi^\alpha\over \dd x^\sigma} dx^\sigma dx^\beta\nonumber\\
   &= \eta_{\alpha\beta} dx^\alpha dx ^\beta + 2 {\dd\xi_\beta\over \dd x^\alpha} dx^\alpha dx^\beta\nonumber\\
   & =\left(\eta_{\alpha\beta} + {\dd\xi_\alpha \over \dd x^\beta} + {\dd \xi_\beta\over \dd x^\alpha} \right)dx^\alpha dx^\beta.
   \end{align}
From equation (\ref{19}), we therefore seek the solution to the set of equations 
   \beq
  \left(  {\dd\xi_\alpha \over \dd x^\beta} + {\dd \xi_\beta\over \dd x^\alpha} \right)dx^\alpha dx^\beta = \left(2r_g\over r_0\right) \left[
 \left(1 -{z\over r_0}\right) (dt^2 +dz^2)  +{2dz \over r_0} (xdx+ydy)\right].
 \eeq
These are:
\beq
{\dd\xi_t\over \dd t} ={\dd\xi_z\over \dd z} = \left( r_g\over r_0\right)\left (1-{z\over r_0}\right),
\eeq
\beq
{\dd\xi_z\over \dd x} + {\dd\xi_x\over \dd z} = {2r_g x\over r_0^2},
\eeq
\beq
{\dd\xi_z\over \dd y} + {\dd\xi_y\over \dd z} = {2r_g y\over r_0^2},
\eeq
\beq
{\dd\xi_t\over \dd z}+{\dd\xi_z\over\dd t}=0,
\eeq
\beq
{\dd\xi_t\over \dd x}+{\dd\xi_x\over\dd t}=0,
\eeq
\beq
{\dd\xi_t\over \dd y}+{\dd\xi_y\over\dd t}=0,
\eeq
\beq
{\dd\xi_x\over \dd y} + {\dd\xi_y\over \dd x} = 0,
\eeq
\beq
{\dd \xi_x\over \dd x} = {\dd \xi_y\over \dd y} = 0.
\eeq

{The simplest solutions} are those with $\xi_x = \xi_y=0$, which we may always choose by rotational symmetry.   The needed equations then reduce to
\beq\label{30}
{\dd\xi_t\over \dd t} ={\dd\xi_z\over \dd z} = \left( r_g\over r_0\right)\left (1-{z\over r_0}\right),
\eeq
\beq\label{31}
{\dd\xi_z\over \dd x} = {2r_g x\over r_0^2},
\eeq
\beq\label{32}
{\dd\xi_z\over \dd y}  = {2r_g y\over r_0^2},
\eeq
\beq\label{33}
{\dd\xi_t\over \dd z}+{\dd\xi_z\over\dd t}=0.
\eeq
\beq\label{34}
{\dd\xi_t\over \dd x}={\dd\xi_t\over \dd y}=0
\eeq
Equation (\ref{30}) has the solutions
\beq
\xi_t =  \left( r_g\over r_0\right)\left (1-{z\over r_0}\right)t +\bar\xi_t(x,y,z), \quad \xi_z = \left( r_g z \over r_0\right)\left (1-{z\over 2r_0}\right)+\bar\xi_z(x,y,t),
\eeq
where $\bar\xi_t$ and $\bar\xi_z$ are arbitrary functions of the indicated variables.  Equations (\ref{31}) and (\ref{32}) may be solved by setting
\beq
\bar\xi_z =\left(r_g\over r_0^2\right) (x^2 +y^2) +\tilde\xi_z(t),
\eeq
where $\tilde\xi_z$ is a function of $t$ only.   We set $\bar\xi_t=0$ to comply with equation (\ref{34}) and choose the zero of time to be independent of position.  Then, equation (\ref{33}) becomes
\beq
-{r_gt\over r_0^2} + {d\tilde \xi_z\over dt} = 0 \rightarrow \tilde\xi_z = {r_gt^2\over 2r_0^2} ,
\eeq
where we have set the unimportant integration constant to zero.   The final form of our $\xi_\mu$ transformation functions is then
\beq
\xi_t = \left( r_g\over r_0\right)\left (1-{z\over r_0}\right)t \simeq {r_g t\over r_0+z},
\eeq
\beq
 \xi_z = \left( r_g z \over r_0\right)\left (1-{z\over 2r_0}\right)+\left(r_g\over r_0^2\right) (x^2 +y^2)+ {r_gt^2\over 2r_0^2},
\eeq
and $\xi_x = \xi_y=0$.   It is helpful at this stage to write $\xi_z$ in physical units.   Denoting the local Newtonian acceleration by $g\equiv GM/r_0^2$, we have
\beq
 \xi_z = g\left( zr_0 \over c^2\right)\left (1-{z\over 2r_0}\right)+\left(g\over c^2\right) (x^2 +y^2)+ {gt^2\over 2} .
\eeq

\begin{figure}
\begin{minipage}[m] {.85\linewidth}
\centering
\includegraphics [width = .85\linewidth]{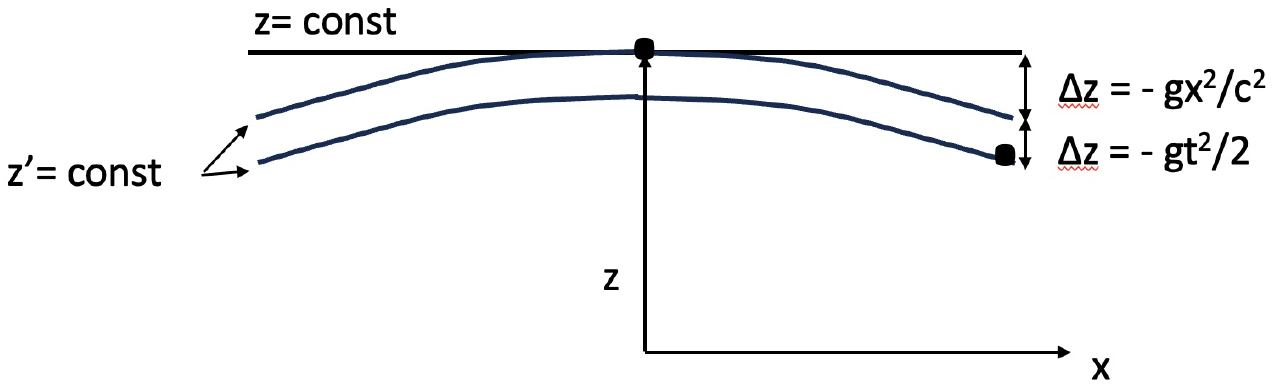}
\caption{\footnotesize Constant $z$ and $z'$ surfaces in the $y=0$ plane.  A test particle, either nonrelativistic or relativistic, moves within a curve of constant $z'$ parallel to the $x$-axis, as the same $z'$ surface falls with {constant acceleration $g$.}  A constant $z'$ surface is shown at initial time zero (upper curve) and a later time $t$ (lower curve).  For a nonrelativistic particle, the curvature of $z'$ is unimportant, and the particle would slide downward very close to the $z$ axis.    This simplification fails as soon as the relativistic regime {is broached.} For a highly relativistic particle (the black dot shown at time zero and time $t$), the curvature triples the local vertical deflection: relativistic deflection occurs on a scale $x=ct$, over which the lowest order spatial curvature matters, {\it even locally,} as $t$ becomes arbitrarily small.   }
\end{minipage}
\end{figure}

For the problem of interest, $z\sim gt^2$, and the final two terms dominate.  We then have
\beq
z' \simeq z +g\left( {t^2\over 2} +{H^2\over c^2} \right),
\eeq
where $H^2=x^2+y^2$ is now the {\it cylindrical} radius (a consistent generalisation of our earlier, rectilinear $x$-based definition of $H$).    The $gt^2/2$ term is familiar from elementary treatments in which the inertial coordinates are linked to a physical, freely-falling observer.   But the spatial transformation (the curvature, in fact) associated with the additional term on the right is $gH^2/c^2= g\beta^2t^2$ for test particles, which becomes important for relativistic trajectories.   In other words, at a given time $t$, a constant $z'$ test particle finds itself at the $z$ location 
\beq
z \simeq {z'} - {gt^2\over 2} - {gH^2\over c^2} = {z'} - {gt^2\over 2}(1+2\beta^2) = {z'} - {gt^2\over 2} \left(3 - {2\over\gamma^2}\right),
\eeq
{which is precisely equation (\ref{morb}).   }

Fig.\ [2] shows an edge-on view of a constant $z$ and $z'$ surfaces in the $y=0$ plane.   As time $t$ increases, a test particle, either relativistic or nonrelativistic, {\it moves within a constant $z'$ surface, which is itself falling by an accumulated amount $-gt^2/2$,} like an ant crawling on the surface toward the perimeter of a downward accelerating, curved parachute.    For a nonrelativistic particle, the spatial curvature is unimportant, as the displacement $x$ remains small on the scale at which curvature would manifest, i.e., $x \ll ct$.  But for relativistic deflections, the $z'$ {surface curvature} is an order unity effect:  it can never be ignored, even in what seems to be a local calculation.   For a photon, the curvature effect is twice the size of the naive ``$gt^2/2$'' EP term, has the same sign, and combines to give a factor of three difference.   Simply by following the distortion of space, photons would be deflected by an angle twice as large as the Newtonian prediction, even in the absence of the ``falling down'' effects of Newtonian gravity.    This latter case may be continuously recovered by finite-mass particles transitioning from relativistic to nonrelativistic velocities.

 \section {Discussion}
In \S 2, the photon and nonrelativistic pebble trajectories can be made the same only by formally setting the curvature parameter $\alpha=0$.  Without spatial curvature, weak gravity is simply a Newtonian force, which produces local accelerations, whose effects are therefore recoverable via kinematic arguments.   The gravitational redshift term gives rise to its eponymous effect as well as to Newtonian gravity in the dynamical equations of motion.   Were we dealing with a nonrelativistic pebble as opposed to a photon, there would be complete consistency among all the approaches, deflections and redshift calculations alike, in essence because $c dt \gg dz$. The point, however, is that elevators and photons do {\it not} fall at the same rate in a gravitational field.   This is not really a new observation.  For example, Moreau, Purdie \& Wood\cite{mpw} have performed a detailed global photon deflection calculation taking care to separate what amounts to kinematic and curvature contributions.   In another study\cite{mnr}, the Schwarzschild metric has been directly broken down into constituent terms that the authors identify with curvature and EP physics.    Nevertheless, as noted by example in the Introduction, the failure of a direct application of the EP based on frames of reference alone seems not to be widely appreciated.    

The factor of three discrepancy for photons we have found in a local calculation arises from the same spatial curvature metric terms that result in the factor of two error in Einstein's first 1911 {\it global} calculation\cite{ae} of the deflection of starlight passing close to the Sun.   The problem is that relativistic orbits require treating the $B$ and $A$ metric coefficients of {equation (\ref{AB})} (i.e., the metric tensor components $g_{tt}$ and $g_{rr}$) on an equal footing, whereas only the $B$ coefficient enters into the calculation for Newtonian orbital dynamics. Similarly, the (vertical) gravitational redshift of a photon requires only the $B$ coefficient in its calculation, and therefore a kinematical approach here is relativistically correct to leading order.   

More generally, one must be careful not to confuse reference frames and coordinates.   While it is true that Einstein's profound initial insight---that coordinates must play a key role in a theory of gravity---came from his realisation that gravity disappears in the {\em reference frame} of a freely-falling observer, in its final manifestation in general relativity the EP is actually best understood as a statement about the mathematical existence of locally flat coordinates, not physical frames of reference\cite{dls}.

Pedagogically, it is at best very misleading to present a kinematic argument for the local deflection of light that captures only one-third of the contribution to the actual displacement.  
Yet more interestingly, the simple calculation discussed here makes it clear that, in principle, a local experiment can be devised 
to distinguish whether an observer is in a Minkowski space or in a freely falling inertial frame.   There is an absolute lower bound to the relativistic particle local deflection angle which is three times the na\"ive EP value of $gt^2/2$.   Semantically, one might argue that the local relativistic deflection is not truly ``local'' in that it involves spatial curvature, even at leading order.   From this viewpoint, embedded in their own constant $z'$ surfaces, photons are indeed dragged downward at the kinematic rate of acceleration, $g$.  This should not, however, obscure the measurable physics.   Equation (\ref{morb}) speaks for itself,  
but it must be a direct measurement in the sense that the {\it difference} between photon and finite mass deflection angles continues to obey the kinematic EP prediction, even in a fully relativistic formalism.     To see this clearly, first rewrite equation (\ref{morb}) as 
\beq 
\Delta z_{\rm m}  =  -{r_g\delta^2 \over 2 }\left( {1\over \beta^2}  +2 \right) =  -{g H^2 \over 2c^2 }\left( {1\over \beta^2}  +2 \right)
\eeq
and therefore (with $\alpha=1$ in equation [\ref{orb}]):
\beq
\Delta z_{\rm m} - \Delta z_{\rm ph} = -{g H^2 \over 2c^2 }\left( {1\over \beta^2} -1 \right)
\eeq
which is identical to the result found in a strictly kinematic approach to the EP.     Even with sophisticated contemporary methods,  neither equation (\ref{orb}) or (\ref{morb}) is amenable to direct measurement, and their difference will not tell us what we need to know.   But then, this class of light deflection measurement is no longer at the cutting edge of experimental general relativity.    The value of the calculation presented here lies in the conceptual clarity afforded by a rather simple, but mathematically explicit, example of the subtlety of a key foundational principle of relativistic gravity theory.   Further discussion on the foundational principles of general relativity discussed in this paper, as well as the mathematics associated with the transformation to local inertial coordinates, may be found in the texts of M{\o}ller\cite{mo}, and Romano and Furnari\cite{rf}.

\begin{acknowledgments}
It is a pleasure to acknowledge helpful conversations with R.\ Blandford, S.\ Blundell, P.\ Ferreira, L.\ Fraser-Taliente, C.\ Gammie, J.\  March-Russell and A.\ Mummery.  
Detailed comments from the anonymous referees significantly improved the presentation.    
\end{acknowledgments}

\end{document}